\documentclass[10pt,aps,twocolumn,prd,noshowpacs,nofootinbib,noshowkeys,floatfix]{revtex4}
\usepackage[dvips]{graphics,graphicx}
\usepackage[colorlinks=true,linktocpage=true,linkcolor=blue,citecolor=blue]{hyperref}
\usepackage[usenames,dvipsnames]{color}
\usepackage{amsmath, amssymb}
\usepackage{multirow}
\usepackage{longtable}
\usepackage{color}
\usepackage[normalem]{ulem}  

\newcommand{\Aparm}{\mathcal{R}(\xi)}

\renewcommand\sout{\bgroup \color{blue} \ULdepth=-.5ex \ULset}

\begin{document}

\preprint{}

\title{Metric anisotropies and emergent anisotropic hydrodynamics}

\author{Ashutosh Dash}
\affiliation{School of Physical Sciences, National Institute of Science Education and Research, HBNI, Jatni-752050, Odisha, India}
\author{Amaresh Jaiswal}
\affiliation{School of Physical Sciences, National Institute of Science Education and Research, HBNI, Jatni-752050, Odisha, India}

\date{\today}

\begin{abstract}

Expansion of a locally equilibrated fluid is considered in an anisotropic space-time given by Bianchi type I metric. Starting from isotropic equilibrium phase-space distribution function in the local rest frame, we obtain expressions for components of the energy-momentum tensor and conserved current, such as number density, energy density and pressure components. In the case of an axis-symmetric Bianchi type I metric, we show that they are identical to that obtained within the setup of anisotropic hydrodynamics. We further consider the case when Bianchi type I metric is a vacuum solution of Einstein equation: the Kasner metric. For axis-symmetric Kasner metric, we discuss the implications of our results in the context of anisotropic hydrodynamics.

\end{abstract}



\maketitle

\section{Introduction}

The success of relativistic hydrodynamics in explaining the space-time evolution of strongly interacting hot and dense matter, produced in relativistic heavy-ion collisions, has initiated new developments in the theoretical formulation of relativistic viscous hydrodynamics \cite{Jeon:2015dfa, Florkowski:2017olj}. In recent years there have been significant advances in our understanding of the theory of relativistic hydrodynamics and its application to high energy heavy ion collisions at Relativistic Heavy-Ion Collider (RHIC) and the Large Hadron Collider (LHC) \cite{Heinz:2013th, Braun-Munzinger:2015hba, Jaiswal:2016hex}. The formulation of dissipative hydrodynamic equations is achieved by obtaining the long wavelength, low frequency limit of the underlying microscopic dynamics of a system \cite{Eckart:1940zz, Grad, Chapman, Landau_Fluid_Mechanics, Romatschke:2009im, Baier:2007ix}. The traditional derivation of dissipative hydrodynamics from kinetic theory relies on a linearization around an equilibrium distribution function which is isotropic in momentum space \cite{Israel:1979wp, Muronga:2003ta, York:2008rr, Denicol:2010xn, Jaiswal:2012qm, Jaiswal:2013fc, Jaiswal:2013npa}. This amounts to expansion of the underlying microscopic kinetic theory in terms of the inverse Reynolds number and Knudsen number around local equilibrium \cite{Jaiswal:2013vta, Bhalerao:2013pza, Jaiswal:2014isa, Florkowski:2015lra, Betz:2008me, Denicol:2012cn}. This type of expansion may not be accurate in situations where deviations from the local equilibrium and/or space-time gradients are large. 

Recent studies have shown that a phase of quantum chromodynamics (QCD) called the quark gluon plasma (QGP), which is created in relativistic heavy-ion collisions, is not isotropic in momentum space. For instance, at very early times, large pressure anisotropies are created in the center of the fireball for viscosities consistent with experimental observations. Moreover, the level of plasma anisotropy increases as one moves away from the center of the fireball to the peripheral regions of the plasma where the temperature is low \cite{Song:2009gc, Martinez:2012tu}. Such large pressure anisotropies indicate large viscous corrections to the distribution function which is contradictory to the near equilibrium assumption of the formulation of viscous hydrodynamic equations. Furthermore, application of traditional linearized viscous hydrodynamics leads to regions of phase space in which the single particle phase-space distribution function may be negative, which may in turn lead to negative longitudinal pressure \cite{Torrieri:2008ip, Martinez:2009mf, Habich:2014tpa}. Depending on whether one considers early times or colder regions of the plasma, the size of these unphysical regions increases. It is important to note that the single particle phase-space distribution function is used to calculate observable plasma signatures, such as dilepton and photon production/flow, quarkonium suppression, and hadronic spectrum through freeze-out. Therefore, inaccuracies in the distribution function can potentially lead to incorrect estimation of these observables \cite{Bhatt:2010cy, Bhatt:2011kx, Bhatt:2011kr, Bhalerao:2013aha}. 

Because of the aforementioned problems in traditional dissipative hydrodynamics, there was motivation to create an alternative framework that could more accurately capture the far-from-equilibrium dynamics of highly momentum-space anisotropic systems. The framework of anisotropic hydrodynamics has proven to be quite successful in this context \cite{Martinez:2010sc, Florkowski:2010cf, Martinez:2010sd, Ryblewski:2011aq, Nopoush:2015yga, Ryblewski:2012rr, Bazow:2013ifa, Tinti:2013vba, Nopoush:2014pfa, Molnar:2016vvu}; see Ref.~\cite{Strickland:2014pga} for a comprehensive review. Anisotropic hydrodynamics is a non-perturbative approximation of relativistic dissipative hydrodynamics which takes into account the large momentum-space anisotropies generated in relativistic heavy-ion collisions. The motivation for the formulation of anisotropic hydrodynamics is to create a framework that is better suited to deal with such large anisotropies and accurately describes several interesting features such as the early time dynamics of the QGP, dynamics near the transverse edges of the fireball and the possibility of large shear viscosity to entropy density ratio $\eta/s$\footnote{Although phenomenological analyses of experimental data suggests that the average value of $\eta/s$ of QGP is small, lattice QCD predicts relatively large values at high temperature \cite{Nakamura:2004sy}. Also see Ref.~ \cite{Niemi:2012ry} for phenomenological implications of temperature dependent $\eta/s$.}. As a consequence, it allows one to extend the regime of applicability of dissipative hydrodynamics to systems that can be quite far from isotropic local thermal equilibrium.

In this paper, we present an alternate derivation of anisotropic hydrodynamic equations by considering the expansion of a locally equilibrated fluid in an anisotropic space-time given by Bianchi type I metric. Assuming isotropic phase-space distribution function at the initial time in the local rest frame, we obtain expressions for components of the energy-momentum tensor and conserved current, such as number density, energy density and pressure components. We show that these expressions are identical to that obtained within the setup of anisotropic hydrodynamics when one considers axis-symmetric Bianchi type I metric.

We further consider the case when Bianchi type I metric is a solution of Einstein equation: the Kasner metric. The Kasner metric describes a curved space-time in general. However, it has been shown that the Kasner space-time can be treated as a well-controlled approximation of a local rest frame of an anisotropically expanding fluid in Minkowski space-time \cite{Sin:2006pv}. Therefore, the Kasner space-time provides a useful framework for studying the anisotropic expansion because of the simplification of the hydrodynamic equations \cite{Culetu:2009xm}. For axis-symmetric Kasner metric, we further discuss the implications of our results in the context of anisotropic hydrodynamics.


\section{The metric}

The most general anisotropic Bianchi type I metric is \cite{Misner:1967uu,Jacobs:1968zz,Thorne:1967zz}
\begin{equation}\label{eq:geLineElement}
ds^2=dt^2-g_{ij}dx^idx^j
\end{equation}
When there is no a priori preferred direction the metric simply takes a diagonal form given as
\begin{equation}\label{eq:LineElement}
ds^2=dt^2 - A^2(t)dx^2 - B^2(t)dy^2 - C^2(t)dz^2.
\end{equation}
The quantities $A(t)$, $B(t)$, and $C(t)$ are scale factors for the expansion along $x$, $y$, and $z$ axes. The metric tensor is diagonal and is given by
\begin{equation}\label{eq:metric}
g_{\mu\nu}={\rm diag}\left[1,-A^2(t),-B^2(t),-C^2(t)\right],
\end{equation}
and the inverse of the metric tensor is given by
\begin{equation}\label{eq:metricinv}
g^{\mu\nu}={\rm diag}\left[1,-\frac{1}{A^2(t)},-\frac{1}{B^2(t)},-\frac{1}{C^2(t)}\right].
\end{equation}
Later, we will specialize to the axis-symmetric case where we will have $A(t)=B(t)$. We will also consider the case when Bianchi type I metric is a solution of Einstein equation: the Kasner metric.


\section{Stress-energy tensor for a gas in thermal equilibrium}

First we will investigate the form of stress energy for a gas of strongly interacting massless particles in thermal equilibrium at a time $t=t_0$. This is possible if the characteristic interaction time is much shorter than the dynamic expansion time of the system.

The stress energy tensor is defined as
\begin{eqnarray}\label{eq:SET}
T^{\mu\nu}&=&\int\!\!\sqrt{-g}\, \frac{d^3p}{p^0}\,p^\mu\,p^\nu\,f\left(x_\mu,p_\mu\right)\\\nonumber
&=&\int\!\!\sqrt{-g}\, \frac{dp^1 dp^2 dp^3}{p^0}\,p^\mu\,p^\nu\,f\left(x_\mu,p_\mu\right),
\end{eqnarray}
where $f\left(x_\mu,p_\mu\right)$ is the scalar distribution function in the relativistic phase space and $g$ is the determinant of metric tensor $g_{\mu\nu}$ and is equal to 
\begin{equation}
g = -A^2B^2C^2 = -V^2
\end{equation}
where $V$ is the physical volume occupied by the particles. Similarly one can define the number density to be,
\begin{equation}\label{eq:NumberDen}
n=\int\!\!\sqrt{-g}\, d^3p f\left(x_\mu,p_\mu\right).
\end{equation}

For ultra-relativistic particles whose masses can be ignored, we know that
\begin{eqnarray}\label{eq:MomentumRelation}
g^{\mu\nu}p_{\mu}p_{\nu}=m^2=0.
\end{eqnarray}
where $m$ is mass. Hence we can write (\ref{eq:MomentumRelation}) as
\begin{equation}\label{eq:EnergyRelation1}
E_0=p_0|_{t=t_0}=\left[\left(\!\frac{p_1}{A(t_0)}\!\right)^{\!2}\! + \left(\!\frac{p_2}{B(t_0)}\!\right)^{\!2}\! + \left(\!\frac{p_3}{C(t_0)}\!\right)^{\!2}\right]^{1/2}
\end{equation}
where we have denoted $E_0$ as the energy of the particles at time $t=t_0$. Since we have thermal equilibrium at time $t=t_0$ we can recast our momenta in spherical polar coordinates $\left(p_0,\theta,\phi\right)$ to extract the components of stress energy tensor ($T_{\mu\nu}$) using Eq.~(\ref{eq:SET}).

The physical components of four-momenta which a local homogeneous observer reads, are defined as
\begin{equation}\label{eq:PhyMomenta}
P_{\mu}={\left(g^{\mu\mu}\right)}^{1/2}p_{\mu}
\end{equation}
such that $P_{\mu}P^{\mu}=m^2$ and no sum is implied in Eq.~(\ref{eq:PhyMomenta}). Thus, in spherical polar coordinates one finds
\begin{eqnarray}\label{eq:FourSpherical}
\nonumber 
P_1&=&\frac{p_1}{A(t_0)}=p_0\sin\theta\,\sin\phi\\
P_2&=&\frac{p_2}{B(t_0)}=p_0\sin\theta\,\cos\phi\\\nonumber
P_3&=&\frac{p_3}{C(t_0)}=p_0\cos\theta
\end{eqnarray}
One can readily verify that the above system of equations satisfy Eq.~(\ref{eq:EnergyRelation1}). We can calculate
the transformation Jacobian of Eq.~(\ref{eq:SET}) to be
\begin{equation}\label{eq:Jacobian}
dp^1 dp^2 dp^3 = \frac{1}{V} p^2_0 dp_0 d\Omega
\end{equation}
We can choose the scalar distribution function $f(x,p)$ as
\begin{equation}\label{eq:DistFunc}
f(x,p) = g_0\frac{1}{e^{E_0/T_0} + r}
\end{equation}
where $g_0$ is the degeneracy factor, $T_0$ is temperature at time $t=t_0$ and $r=0,~+1$ and $-1$ for Boltzmann, Fermi-Dirac and Bose-Einstein distribution functions, respectively. Note that we have assumed Boltzmann's constant and Planck's constant to be unity, i.e., $k=\hbar=1$.

Inserting Eqs.~(\ref{eq:FourSpherical})-(\ref{eq:DistFunc}) into Eqs.~(\ref{eq:SET}) and (\ref{eq:NumberDen}) and doing the angular
integration, one easily obtains the following equilibrium relations:
\begin{eqnarray}\label{eq:EqullibrmSET}
n &=& g_1 (T_0)^3, \nonumber\\
\varepsilon &=& T_{00} = g_2 (T_0)^4, \\ 
{\cal P} &=& T_{11} = T_{22} = T_{33} = \frac{1}{3}\varepsilon, \nonumber
\end{eqnarray}
where we have absorbed some constants appearing after integration into the redefined degeneracy factors $g_1$ and $g_2$. Thus, we have established that the form of the stress energy tensor completely agrees with that of a gas in thermal equilibrium with its surrounding having temperature $T=T_0$. In the next section we will derive the evolution of stress energy when the gas gets completely decoupled from its surrounding.


\section{ The collision-less stress-energy tensor}

We idealize the decoupling of the gas from its surrounding happens at time $t=t_0$, such that after time $t_0$ the gas experiences a collision-less adiabatic expansion or contraction as specified by the metric of Eq.~(\ref{eq:LineElement}). Also, Liouville's theorem guarantees that the distribution function $f(x,p)$ of Eq.~(\ref{eq:DistFunc}) remains constant throughout the phase space for all time during the evolution. This in turn implies that the energy $E$ and the temperature $T$, at a given time $t$, are red-shifted by the same amount, i.e.,
\begin{equation}\label{eq:RedShift}
\frac{E}{E_0}=\frac{T}{T_0}=z
\end{equation}
where $z$ is the usual red-shift factor. 

The evolution of the stress-energy tensor depends only on the function $z$ which needs to be determined. From Eq.~(\ref{eq:MomentumRelation}) the energy $E$ for the particles evolving by the metric given in Eq.~(\ref{eq:LineElement}) at time $t$ is
\begin{equation}\label{eq:EnergyRelation}
E_0=\left[\left(\frac{p_1}{A(t_0)}\right)^2+\left(\frac{p_2}{B(t_0)}\right)^2+\left(\frac{p_3}{C(t_0)}\right)^2\right]^{1/2}.
\end{equation}
Since, the 3-momenta $p_i$ are constants of motion, i.e., $dp_i/d\tau=0$ (as shown in Appendix A), the contra-variant components of Eq.~(\ref{eq:EnergyRelation}) are:
\begin{equation}\label{eq:EnergyRelation2}
E_0=\left[\left(\frac{A^2(t)p^1}{A(t_0)}\right)^2+\left(\frac{B^2(t)p^2}{B(t_0)}\right)^2+\left(\frac{C^2(t)p^3}{C(t_0)}\right)^2\right]^{1/2}.
\end{equation}
Now, using Eqs.~(\ref{eq:PhyMomenta}) and (\ref{eq:FourSpherical}) in Eq.~(\ref{eq:EnergyRelation2}), we can easily find the red-shift factor $z$ to be,
\begin{align}\label{eq:MasterEqn}
z=\Bigg[&{\left(\frac{A(t)\sin\theta\,\sin\phi}{A(t_0)}\right)}^2+{\left(\frac{B(t)\sin\theta\,\cos\phi}{B(t_0)}\right)}^2 \nonumber\\
&+{\left(\frac{C(t)\cos\theta}{C(t_0)}\right)}^2\Bigg]^{-1/2}
\end{align}
From Eq.~($\ref{eq:MasterEqn}$), we see that the characteristic temperature is dependent on the direction of motion of particles.

Using Eqs.~(\ref{eq:RedShift}) and (\ref{eq:MasterEqn}), we can calculate the components of the stress-energy tensor at a later time $t>t_0$ from Eq.~(\ref{eq:EqullibrmSET}). We proceed in the same way as before except the angular integration is altered. We obtain
\begin{align}
n&=\frac{n_0}{4\pi}\int_0^{2\pi}d\phi\int_0^{\pi}\sin\theta z^3d\theta \label{eq:ndensity}\\
\varepsilon&=T_{00}=\frac{{\varepsilon}_0}{4\pi}\int_0^{2\pi}d\phi\int_0^{\pi}\sin\theta z^4d\theta, \label{eq:edensity}\\
{\cal P}_x&=T_{11}=\frac{{3({\cal P}_x)}_0}{4\pi}\int_0^{2\pi}\cos^2\phi~d\phi\int_0^{\pi}\sin^3\theta z^4d\theta, \label{eq:Pxdensity}\\
{\cal P}_y&=T_{22}=\frac{{3({\cal P}_y)}_0}{4\pi}\int_0^{2\pi}\sin^2\phi~d\phi\int_0^{\pi}\sin^3\theta z^4d\theta, \label{eq:Pydensity}\\
{\cal P}_z&=T_{33}=\frac{{3({\cal P}_z)}_0}{4\pi}\int_0^{2\pi} d\phi\int_0^{\pi}\sin\theta\cos^2\theta z^4d\theta. \label{eq:Pzdensity}
\end{align}
If we assume an axis-symmetric case, i.e $\frac{A(t)}{A(t_0)}=\frac{B(t)}{B(t_0)}=\xi_1$ and $\frac{C(t)}{C(t_0)}=\xi_3$ in which case the system of equations (\ref{eq:ndensity})-(\ref{eq:Pydensity}) reduce to a more tractable form:
\begin{align}
n&=\frac{n_0}{2}\int_{-1}^{1}{\left(\lambda^2(\xi_3^2-\xi_1^2)+\xi_1^2\right)}^{-3/2}d\lambda, \label{eq:nbdensity}\\
\varepsilon&=\frac{\varepsilon_0}{2}\int_{-1}^{1}{\left(\lambda^2(\xi_3^2-\xi_1^2)+\xi_1^2\right)}^{-2}d\lambda, \label{eq:endensity}\\
{\cal P}_{\bot}&=\frac{{3(P_{\bot})}_0}{4}\int_{-1}^{1}(1-\lambda^2){\left(\lambda^2(\xi_3^2-\xi_1^2)+\xi_1^2\right)}^{-2}d\lambda, \label{eq:Pplldensity}\\
{\cal P}_{\parallel}&=\frac{{3(P_{\parallel})}_0}{2}\int_{-1}^{1}\lambda^2{\left(\lambda^2(\xi_3^2-\xi_1^2)+\xi_1^2\right)}^{-2}d\lambda,\label{eq:Pperdensity}
\end{align}
where $\lambda=\cos\theta$, we have defined ${\cal P}_x={\cal P}_y={\cal P}_{\bot}$ and ${\cal P}_z={\cal P}_{\parallel}$.

Integrating Eqs.~(\ref{eq:nbdensity})-(\ref{eq:Pperdensity}), we get
\begin{align}
n&=\frac{n_0}{\xi_1^3\xi^{1/2}},\label{eq:NumDensity}\\
\varepsilon&=\frac{\varepsilon_0\xi^{-1}}{2\xi_1^4}\left(1+\frac{\xi\arctan\sqrt{\xi-1}}{\sqrt{\xi-1}}\right)=\frac{\varepsilon_0\Aparm}{\xi_1^4},\label{eq:EnDensity}\\
{\cal P}_{\bot}&=\frac{{3({\cal P}_{\bot})}_0}{4\xi_1^4}\left(\frac{1}{\xi-1}+\frac{(\xi-2)\arctan\sqrt{\xi-1}}{{(\xi-1)}^{3/2}}\right) \nonumber\\
&=\frac{{3({\cal P}_{\bot})}_0}{2\xi_1^4}\left(\frac{1+\xi(\xi-2)\Aparm}{\xi(\xi-1)}\right), \label{eq:LongPressure}\\
{\cal P}_{\parallel}&=\frac{{3({\cal P}_{\parallel})}_0}{2\xi_1^4}\left(\frac{1}{\xi(1-\xi)}+\frac{\arctan\sqrt{\xi-1}}{{(\xi-1)}^{3/2}}\right) \nonumber\\
&=\frac{{3({\cal P}_{\parallel})}_0}{\xi_1^4}\left(\frac{\xi\Aparm-1}{\xi(\xi-1)}\right), \label{eq:LongPressure}
\end{align}
where $\xi=\frac{\xi_3^2}{\xi_1^2}$ and $\Aparm=\frac{1}{2\xi}\left(1+\frac{\xi\arctan\sqrt{\xi-1}}{\sqrt{\xi-1}}\right)$ for $\xi>1$, while we substitute $\xi$ as $\frac{1}{\xi}$ for $\xi<1$.


\section{The collision-less Boltzmann equation}

The expression for components of the energy-momentum tensor and conserved current, given in Eqs.~(\ref{eq:NumDensity})-(\ref{eq:LongPressure}), for a system of anisotropically expanding collision-less plasma, can also be obtained by considering the Boltzmann equation in the free streaming case. The collision-less Boltzmann equation for the Bianchi type I metric is given as:
\begin{equation}
p^{\mu}\partial_{\mu}f-\Gamma^{\mu}_{\alpha\beta}p^{\alpha}p^{\beta}\frac{\partial f}{\partial p^{\mu}}=0,
\end{equation}
Taking into account that $f$ cannot depend on the position $x_i$, because of homogeneity of space, the collision-less Boltzmann equation thus becomes:
\begin{eqnarray}
    p^{0}\partial_{0}f-\Gamma^{\mu}_{\alpha\beta}p^{\alpha}p^{\beta}\frac{\partial f}{\partial p^{\mu}}=0,\\
     p^{0}\partial_{0}f-\left(2\Gamma^{\mu}_{0j}p^{0}+\Gamma^{\mu}_{ij}p^{i}\right)p^{j}\frac{\partial f}{\partial p^{\mu}}=0
\end{eqnarray}
The non-zero components of the Christoffel symbols are $\Gamma^x_{0x}=\frac{\dot{A}}{A}$,  $\Gamma^y_{0y}=\frac{\dot{B}}{B}$ and $\Gamma^z_{0z}=\frac{\dot{C}}{C}$. Substituting
above gives us:
\begin{equation}
     \partial_{0}f- 2\left(\frac{\dot{A}}{A} p^{x}\frac{\partial f}{\partial p^{x}}+\frac{\dot{B}}{B} p^{y}\frac{\partial f}{\partial p^{y}}+ \frac{\dot{C}}{C} p^{z}\frac{\partial f}{\partial p^{z}}\right)=0
\end{equation}
Solving the above equation by the method of characteristics yields $f= f(A^2(t)p^x,B^2(t)p^y,C^2(t)p^z)$ $=f(A(t)P^x,B(t)P^y,C(t)P^z)$, where we have used Eq.~(\ref{eq:PhyMomenta}) and redefined momenta in terms of physical momenta \cite{Romatschke:2015dha}.
 
From the above equation we see that for a collision-less system, the momentum dependence of the distribution function should be of the form $f=f(A^2(t)p^x,B^2(t)p^y,C^2(t)p^z)$. Using this form of functional dependence in the equilibrium distribution function, given in Eq.~(\ref{eq:DistFunc}) and taking the appropriate moments again leads to the same expressions for the components of the energy-momentum tensor and conserved current, Eqs.~(\ref{eq:NumDensity})-(\ref{eq:LongPressure}), of a system of anisotropically expanding collision-less plasma \cite{Keegan:2015avk}. This however is not surprising because in the previous section we used the fact that the particle momenta, $p_i$, are constants of motion, i.e., particles do not suffer any collision and are free streaming. Therefore the solution of collision-less Boltzmann equation should also lead to the same expressions for the components of the energy-momentum tensor and conserved current, as demonstrated here. We note that anisotropic expansion through metric, as considered here, naturally leads to Romatshke-Strickland form of the distribution function \cite{Romatschke:2003ms}.


\section{The Kasner metric}

We shall restrict ourselves here even further to the classical vacuum solutions of Einstein's equation, and consider only the subclass of Bianchi-I metrics in which the expansion factors take the Kasner form \cite{Belinsky:1970ew,MTW:1970}
\begin{equation}\label{Kasner}
ds^2=dt^2-t^{2a}dx^2-t^{2b}dy^2-t^{2c}dz^2,
\end{equation}
where $a$, $b$ and $c$ are three parameters that are related to each other by the equations
\begin{align}
a + b + c &= 1 \label{eq:Condn1}\\
a^2 + b^2 + c^2 &= 1. \label{eq:Condn2}
\end{align}
The above constraints are obtained by requiring that the metric given in Eq.~(\ref{Kasner}) is a vacuum solution of the Einstein's equation. However, as was shown in Ref.~\cite{Belinsky:1970ew}, the fluid satisfies the above relations if we impose conformal invariance. In general, the Kasner metric describes a curved space-time. However, it was shown that the Kasner space-time can be treated as an approximation of a local rest frame of an anisotropically expanding fluid in Minkowski space-time \cite{Sin:2006pv}. Hence, the hydrodynamic equations for anisotropic expansion takes a simple form in Kasner space-time \cite{Culetu:2009xm}.

Since the particle current must be conserved, the number density $n$ of particles that is measured by a co-moving observer satisfies 
the continuity equation
\begin{equation}\label{eq:Num}
 \frac{dn}{dt}+\Gamma_{i0}^in=0.
\end{equation}
The non-vanishing components of the Christoffel symbols for Kasner metric are,
\begin{equation}\label{eq:Christoffel}
\Gamma_{10}^1=\frac{a}{t},\qquad \Gamma_{20}^2=\frac{b}{t},\qquad
\Gamma_{30}^3=\frac{c}{t}.
\end{equation}
Using Eq.~(\ref{eq:Christoffel}) in Eq.~(\ref{eq:Num}) we have
\begin{eqnarray}
 \frac{dn}{dt}+(a+b+c)\frac{n}{t}&=&0\\
 \frac{dn}{dt}+\frac{n}{t}&=&0\label{NumConsvn}
\end{eqnarray}
where in the second equation we have used Eq.~(\ref{eq:Condn1}). On integrating Eq.~(\ref{NumConsvn}) we have,
\begin{equation}
n=\frac{n_0t_0}{t}.
\end{equation}
It is interesting to note that the above equation holds for all Kasner type expansion. As demonstrated in the following, the Milne metric turns out to be a special case of Kasner metric. 

From Eqs.~(\ref{eq:Condn1}) and (\ref{eq:Condn2}), we see that out of the three parameters $a$, $b$ and $c$, only one is independent. If we impose an additional constraint of azimuthal symmetry, we have only two possibilities for $(a,b,c)$:
\begin{equation}\label{eq:KasnerCond}
   \text{Case I:}~\left(0,0,1\right) \qquad \text{Case II:}~\left(\frac{2}{3},\frac{2}{3},-\frac{1}{3}\right).
\end{equation}
The first one of course reduces to the usual Milne coordinates by a coordinate transformation $t\sinh z=y$ and $t\cosh z=\tau$ where $y$ and $\tau$ are rapidity and proper-time. The second one is a new finding in the context of azimuthally symmetric anisotropic hydrodynamics. It is important to note that for Case I, all the components of the Riemann curvature tensor vanishes and hence can be obtained by a general co-ordinate transformation of the Minkowski metric. On the other hand, Case II has curvature and therefore can not be obtained by a general co-ordinate transformation of the Minkowski metric which is flat. 

Imposing the condition in Eq.~(\ref{eq:KasnerCond}) on the variable $\xi$ gives us
\begin{equation}\label{eq:KasnerTime}
    \text{Case I:}~\xi=\frac{t^2}{t_0^2}\hspace{4em} \text{Case II:}~\xi=\frac{t_0^2}{t^2}
\end{equation}
Case I refers refers to longitudinal expansion while Case II denotes transverse expansion. We note that case I corresponds to the usual free streaming solution in Bjorken expansion which has been obtained in the past by other authors \cite{Kapusta:1992uy, Martinez:2008di, Florkowski:2008ag}. It is also interesting to note that a system which is Bjorken expanding in Minkowski space-time is staic in the Milne co-ordinate system, which is Case I of Eqs.~(\ref{eq:KasnerCond}) and (\ref{eq:KasnerTime}).


\section{Results and Discussion}

In this section, we investigate the evolution of stress-energy tensor of the fireball that has experienced a relative longitudinal contraction or expansion at time $t$ along the $z$ axis such that $t>t_0$ where $t_0$ is the time of isotropization.

\begin{figure}[t]
\begin{center}
\includegraphics[width=\linewidth]{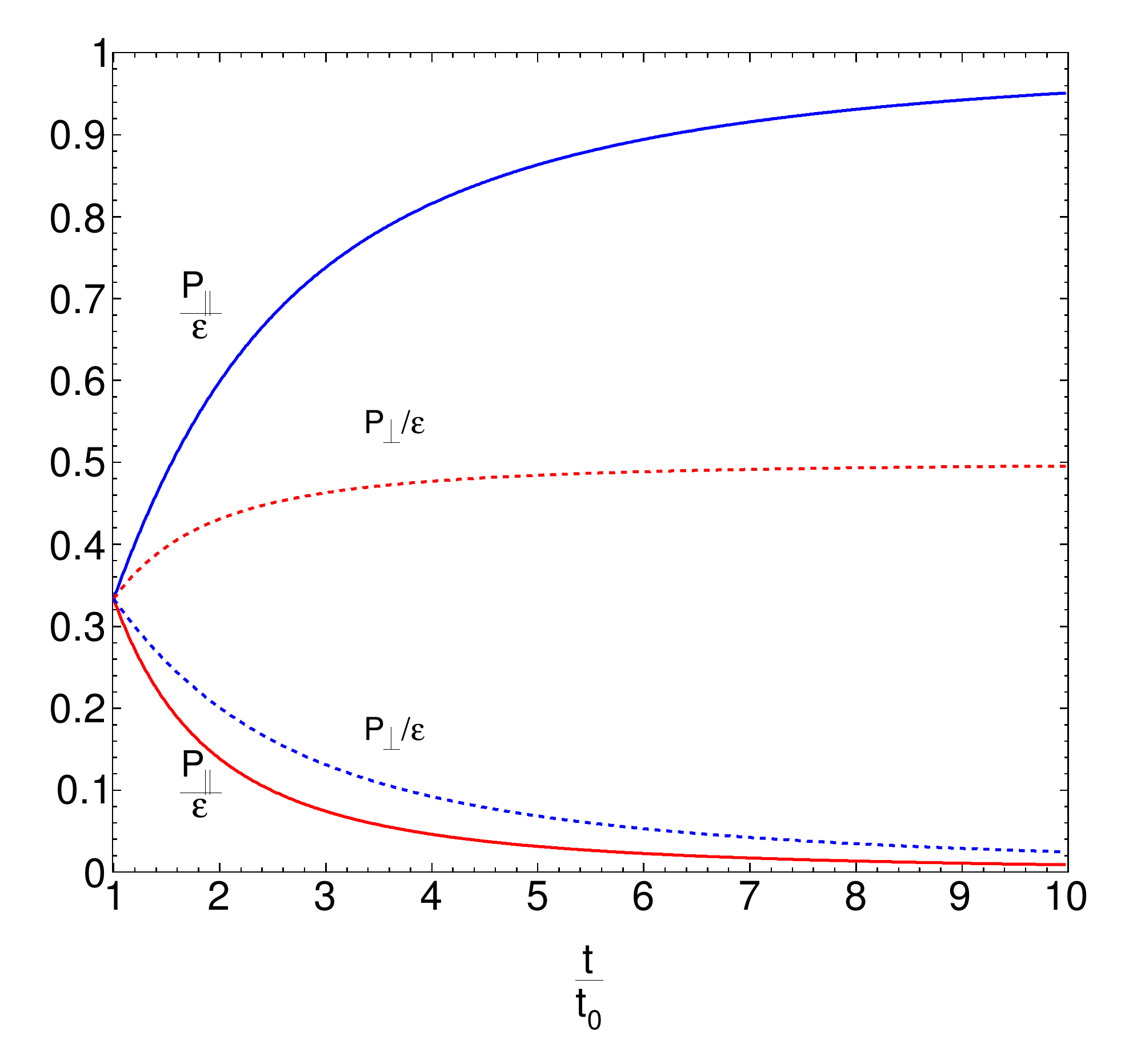}\end{center}
\vspace{-0.5cm}
 \caption{Evolution of longitudinal and transverse pressures, scaled by the energy density, for Case I (red) and Case II (blue) as described in the text.}
\label{fig:figure1}
\end{figure}

For a violent longitudinal expansion $\xi>1$ and the form of $\xi$ in Eq.~(\ref{eq:KasnerTime}) is that of Case I in Fig.~\ref{fig:figure1}. By Eqs.~(\ref{eq:NumDensity})-(\ref{eq:LongPressure}) this induces a stress-energy tensor of the form:
\begin{equation}
{\cal P}_{\parallel}= 0\hspace{4em} {\cal P}_{\bot}=\frac{1}{2}\varepsilon
\end{equation}
as the anisotropic longitudinal expansion increases, i.e., in the limit $1/\xi\rightarrow 0$. Throughout the evolution starting from initial isotropization at time $t_0$ to the final asymptotic limit we have ${\cal P}_{\parallel}<{\cal P}_{\bot}$ as dictated by Eqs.~(\ref{eq:NumDensity})-(\ref{eq:LongPressure}).

The scenario of Case II happens when there is a violent transverse expansion i.e. $\xi<1$ which leads to stress-energy of the limiting form:
\begin{equation}
{\cal P}_{\parallel}= \varepsilon\hspace{4em} {\cal P}_{\bot}=0 
\end{equation}
as the anisotropic transverse expansion increases, i.e., in the limit $\xi\rightarrow 0$. Equations (\ref{eq:NumDensity})-(\ref{eq:LongPressure}) imply that expansion along the transverse direction is accompanied by a simultaneous longitudinal contraction which eventually builds up an enormous pressure along the $z$ direction while the slow expansion along the transverse direction continues, as evident in Fig.~\ref{fig:figure1}. 

In both cases, we observe that the system never reaches isotropic state. This is due to the fact that we have considered free streaming, i.e., non interacting evolution. Note that this is in contrast to dissipative hydrodynamics where the evolution drives the system towards equilibrium.


\section{Summary and outlook}

In this paper, we have considered the free streaming of a locally equilibrated fluid in an anisotropic space-time given by Bianchi type I metric. We obtained expressions for components of the energy-momentum tensor and conserved current, such as energy density, pressure components and number density, for an asymptotic observer. In the case of an axis-symmetric Bianchi type I metric, we showed that they are identical to that obtained within the setup of anisotropic hydrodynamics. We further considered the case when Bianchi type I metric is a solution of Einstein equation: the Kasner metric. For axis-symmetric Kasner metric, we discussed the implications of our results in the context of anisotropic hydrodynamics. 

The framework presented in this paper may also find applications in the context of cosmology. In standard cosmological model, it is assumed that the space-time is isotropic about every point in space and time. However, after the discovery of temperature anisotropies of the Cosmic Microwave Background (CMB), we now know that the universe is isotropic up to small perturbations. If the CMB temperature were isotropic about every point in space-time, then the universe can be described by an exact Friedmann-Lemaitre model \cite{Ehlers:1966ad}. However, since the CMB radiation is not exactly isotropic, it can be described by a perturbed Friedmann-Robertson-Walker metric which can be obtained as a special case of Bianchi type I metric. Since the framework of anisotropic hydrodynamics is well studied, one may apply similar techniques in cosmological models.

Looking forward, it will be interesting to consider an interacting medium within the present setup by considering all possible corrections to the energy momentum tensor up to a particular order in gradients. Alternatively, one can consider an evolving medium through interactions, i.e., in the presence of a non-vanishing collision kernel in the Boltzmann equation. This will lead to viscous corrections in the local distribution function. One can also study the evolution of dissipative quantities within this setup. We leave these questions for future studies.


\begin{acknowledgments}

The authors would like to thank Sourendu Gupta, Bedangadas Mohanty and Victor Roy for helpful comments. A.D. thanks Ananta P. Mishra and A.J. thanks Wojciech Florkowski, Michael Strickland and Leonardo Tinti for useful discussions. A.J. is supported in part by the DST-INSPIRE faculty award under Grant No. DST/INSPIRE/04/2017/000038.

\end{acknowledgments}


\appendix
\setcounter{secnumdepth}{0}

\section{APPENDIX: GEODESIC EQUATION AND FREE STREAMING}
\label{Appendix A}

In this Appendix we show that after the decoupling time $t_0$ as the particles stream freely through space-time, the momenta $p_{\mu}$ of particles are constants of the motion along their phase-space trajectories. Consider the general geodesic equation for the Bianchi type I metric
\begin{equation}\label{eq:Geodesic}
    \frac{dp^{\mu}}{d\tau}+\Gamma^{\mu}_{\rho\sigma}p^{\rho}p^{\sigma}=0
\end{equation}
where $\Gamma^{\mu}_{\rho\sigma}$ are the usual Christoffel  symbols, $p^{\mu}$ is the four momentum of the particle and $\tau$ is the proper time. For convenience we consider only the $x$ component of Eq.~(\ref{eq:Geodesic}) and other components could be derived straightforwardly. For the $x$ component the non-vanishing components of $\Gamma^x_{\rho\sigma}$ are $\Gamma^x_{0x}=\Gamma^x_{x0}=\frac{\dot{A}}{A}$. Substituting this into the geodesic equation, Eq.~(\ref{eq:Geodesic}), we have,
\begin{eqnarray}
 \frac{dp^x}{d\tau}+2\frac{\dot{A}}{A}p^0p^x=0\\
 \frac{dp^x}{d\tau}+\frac{2p^x}{A}\frac{dA}{d\tau}=0\label{eq:Geodesic II}
\end{eqnarray}
where we used the identity $\dot{A}p^0=\frac{dA}{dt}\frac{dt}{d\tau}=\frac{dA}{d\tau}$. We can rewrite Eq.~(\ref{eq:Geodesic II}) as
\begin{equation}
    \frac{d}{d\tau}\left(A^2p^x\right)=\frac{dp_x}{d\tau}=0
\end{equation}
which implies $p_x=const.$ is a constant of motion. Similar relation hold for other components of $p_{\mu}$.

\bibliographystyle{plain}


\end{document}